**Magnetization asymmetry of type-II superconductors in high magnetic fields**

**D M Gokhfeld, D A Balaev, M I Petrov, S I Popkov, K A Shaykhutdinov, V V Valkov**

L V Kirensky Institute of Physics SB RAS, Krasnoyarsk, 660036, Russia

E-mail: gokhfeld@iph.krasn.ru

**Abstract.** Inhomogeneous distribution of the pinning force in superconductor results in a magnetization asymmetry. A model considering the field distribution in superconductor was developed and symmetric and asymmetric magnetization loops of porous and textured $Bi_{1.8}Pb_{0.3}Sr_{1.9}Ca_2Cu_3O_x$ were fitted. It is found that the thermal equilibrium magnetization realizes in crystals smaller than some size depending on temperature and magnetic field.

PACS: 74.25.Ha, 74.25.Wx, 74.81.Bd

## 1. Introduction

Critical characteristics of superconductors may be derived from the dependence of magnetization *M* on magnetic field *H*. However many factors affect the shape of the hysteretic *M*(*H*) curve. The same sample can have a various magnetization loop shape if it has been measured at different temperatures or up to different values of the maximal magnetic field. Also the geometric shape of the sample, its structure and homogeneity influence on the *M*(*H*) dependence [1,2].

At temperatures higher than ~ 10 K the high-field *M*(*H*) dependencies of high-$T_c$ superconductors have a pronounced asymmetry with respect to the $M = 0$ axis, so the main parts of hysteretic loops locate in the second and the fourth quadrants. This asymmetry grows with increasing temperature. Chen et al. [3] following approach [4] considered the total magnetization as the result of both the surface and the volume supercurrents, so the equilibrium magnetization and the surface barrier were included. As suggested in [3], the surface layer of a bulk sample, with a thickness of the order of the London penetration depth, is always in equilibrium. In the remnant volume of the sample the critical state model can be used with modified boundary conditions [3].

Because the model [3] is phenomenological and has a large number of fitting parameters, we were interested to obtain the similar results using a different approach. Earlier, the approach [5] was successfully applied for symmetric hysteretic loops of Tl1223 and Bi2223 measured at low temperature [5,6]. In this article we have reexamined the model [5] and have modified it to fit the asymmetric magnetization loops. The resulted model explains origin of the $M(H)$ asymmetry and gives results consistent with [3]. The asymmetric $M(H)$ dependencies of the porous high-$T_c$ superconductor and textured one with the same chemical composition were fitted by the model. Our approach on the $M(H)$ dependencies is described in Sec. 2. In Sec. 3 the magnetization loops of porous and textured $Bi_{1.8}Pb_{0.3}Sr_{1.9}Ca_2Cu_3O_x$ samples are presented. Discussion and conclusion are given in Sec. 4.

## 2. Model

All realizing $M(H)$ dependencies of superconductors are distributed between two extreme cases: a symmetric magnetization loop and a fully reversible $M(H)$ curve. Symmetry of $M(H)$ curve means that $M^-(H) = -M^+(H)$ at high $H$ except region near maximal applied field $H_m$, where $M^-(H) \leq 0$ is the branch of magnetization curve measured during the field increasing from 0 to $H_m$, $M^+(H)$ is the branch of magnetization curve measured during the field decreasing $H_m$ to 0. Symmetric $M(H)$ dependencies are observed at low temperatures when a strong pinning of the vortices realizes. If there is no pinning in the sample then the $M(H)$ dependence is fully reversible, $M^-(H) = M^+(H)$ for any $H$. The reversible $M(H)$ dependencies are observed for clean superconductors. High-$T_c$ superconductors demonstrate crossover from the symmetric magnetization curves at low $T$ to the reversible $M(H)$ dependencies at high $T$ while pinning force decreases.

The critical state model describes successfully symmetric $M(H)$ dependencies in high magnetic fields. In version [5] it includes the next points:

i. Magnetization of the material, which consists of a large number of cylindrical granules, described by the statistical ensemble $\varphi(R)$, is determined by

$$4\pi M(H) = -H + (1-P)\mu_n H + 2/R_0^2 \int_0^\infty \varphi(R) \int_0^R rB(r)drdR\,, \tag{1}$$

where $P$ is the fraction of the material concentrated in the superconducting granules, $\mu_n$ is the magnetic permeability of the intergranular material, $\varphi(R)$ is the distribution density of the superconducting granules with respect to the radius $R$, $B(r)$ is the magnetic induction in the sample. A homogeneous sample is described by (1) with $P = 0$ and $R_0 = R$.

ii. The distribution of the magnetic induction $B(r)$ in the granule is deduced from the critical state equation $dB(r)/dr = \pm(4\pi/c)j_c(B)$, where $j_c$ is the critical current density. Separate sets of eq-uations (2a,b,c) are resulted for three main parts of the $M(H)$ hysteresis. 1) During the initial field increasing from 0 to $H_m$, the $B(r)$ dependence is determined by equation:

$$\Phi(B) - \Phi(\mu_n H) = -(4\pi/c)\, j_{c0}\, (R-r). \tag{2a}$$

2) For decreasing $H$ from $H_m$ to 0 (i.e. $M^+(H)$ branch), the $B(r)$ dependence consists of two curves which are described by equations:

$$\Phi(B) - \Phi(\mu_n H) = (4\pi/c)\, j_{c0}\, (R-r),$$
$$\Phi(B) - \Phi(\mu_n H_m) = -(4\pi/c)\, j_{c0}\, (R-r). \tag{2b}$$

3) Then, for changing $H$ from 0 to $-H_m$ (i.e. $M^-(H)$ branch), the $B(r)$ dependence consists of three parts:

$$\Phi(-B) - \Phi(-\mu_n H) = -(4\pi/c)\, j_{c0}\, (R-r),$$
$$\Phi(B) + \Phi(-\mu_n H) = (4\pi/c)\, j_{c0}\, (R-r),$$
$$\Phi(B) - \Phi(\mu_n H_m) = -(4\pi/c)\, j_{c0}\, (R-r). \tag{2c}$$

Here $j_{c0}$ is the critical current density at $H = 0$, $\Phi(B)$ is a nondecreasing function, such that $d\Phi(B)/dB = (j_c(B)/j_{c0})^{-1}$ [5].

The critical state model implies the strong pinning of vortices in a sample. In reality some distribution of the pinning force $F_p(r)$ is realized in granules. Therefore a region near granule surface can appear at high $T$ and high $H$ where $F_p$ is such small that vortices are not pinned. So the thermal equilibrium magnetization is produced by the surface supercurrent circulating in this region. This conclusion is coincide with idea [3,4]. To account the equilibrium magnetization we should modify equations 2a,b,c. We introduce new characteristic size $l_s$ determining the depth of surface layer where the vortices are not pinned. Boundary conditions are changed due to the surface supercurrent $j_s$ for the first equations in each set (2a), (2b) and (2c) and equation for the $B(r)$ distribution in the equilibrium surface region is added:

$$\Phi(B) - \Phi(\mu_n H) = -(4\pi/c)\, j_{s0}\, (R-r),$$
$$\Phi(B) - \Phi(B_s) = -(4\pi/c)\, j_{c0}\, (R-r-l_s), \tag{3a}$$

$$\Phi(B) - \Phi(\mu_n H) = -(4\pi/c)\, j_{s0}\, (R-r),$$
$$\Phi(B) - \Phi(B_s) = (4\pi/c)\, j_{c0}\, (R-r-l_s),$$
$$\Phi(B) - \Phi(\mu_n H_m) = -(4\pi/c)\, j_{c0}\, (R-r), \tag{3b}$$

$$\Phi(-B) - \Phi(-\mu_n H) = -(4\pi/c)\, j_{s0}\, (R-r),$$
$$\Phi(-B) - \Phi(|B_s|) = -(4\pi/c)\, j_{c0}\, (R-r-l_s),$$
$$\Phi(B) + \Phi(-\mu_n H) = (4\pi/c)\, j_{c0}\, (R-r),$$
$$\Phi(B) - \Phi(\mu_n H_m) = -(4\pi/c)\, j_{c0}\, (R-r), \tag{3c}$$

here $j_{s0}$ is the surface critical current density at $H = 0$, $B_s$ is the induction value at $r = R–l_s$. The separate notation for $j_s$ is introduced to conformity with approach [3]. However we suppose that in most cases $j_s = j_c$.

Figure 1 displays the computed $B(r)$ dependencies after the field decreasing from $H_m$ to $0.6H_m$. Curve 1 on figure 1 (solid line) is the $B(r)$ dependence for $l_s = 0$, curve 2 (circles) is the $B(r)$ dependence for $0 \le l_s \le R$, curve 3 on figure 1 (dotted line) is the $B(r)$ dependence for $l_s \ge R$. The $B(r)$ distributions in figure 1 were calculated with the $j_c(B)$ dependence characterized by qualitatively different behaviors in scales of low and high fields [5]:

$$j_{\mathrm{c}}(B) = \frac{j_{\mathrm{c}0}}{\frac{1+QB/B_1}{1+B/B_1} + \left(\frac{B}{B_0}\right)^g}, \tag{4}$$

where $j_{c0}$ is the critical current density at $B = 0$, $B_1$ and $B_0$ are the parameters with the induction measure which determine the characteristic scales. Dimensionless parameters $Q > 0$ and $g > 0$ determine the $j_c$ decrease rate. Such $j_c(B)$ dependence is convenient because it is easy integrated.

Some $M(H)$ dependencies computed by the obtained model are presented in figures 2 and 3. This model is strongly adequate for the $M(H)$ dependencies in high fields only. The surface barrier should be account for detailed calculation of low-field magnetization as it was realized in [3]. In contrast to the model [3], the presented model accounts $l_s$, the position of the boundary between the equilibrium layer and the central region with strong pinning in the granules.

The size of region, where pinning is missing, $l_s$, depends on $H$ and $T$. If the condition $l_s \ge R$ takes place for granules, the width of the magnetization loop $|M^-(H) - M^+(H)|$ is equal to 0. The crossover from the symmetric magnetization to the reversible $M(H)$ dependence can be observed while $l_s$ changes from 0 to $l_s \ge R$. The value of $l_s$ is believed to be about the London penetration depth $\lambda$ [3]. Therefore if a superconducting crystal or grains in a polycrystalline sample have a size smaller than $2\lambda$ then pinning is expected to be missing in all volume and the $M(H)$ dependencies are reversible at any temperature.

## 3. Experiment

Synthesis of a porous $Bi_{1.8}Pb_{0.3}Sr_{1.9}Ca_2Cu_3O_x$ (Bi2223) is described in [7]. This material is formed by disordered and unpacked flakes-like crystallites with the width of ~1 μm and the lengthwise sizes of ~20 μm. The density of the porous Bi2223 is equal to 1.6 g/cm$^3$. The temperature of superconducting transition is 113 K.

The textured Bi2223 was prepared from the porous one [8]. The textured ceramics with the texture degree of 0.97 consists of ordered crystallites with the sizes of ~1 μm along the *c* axis and ~10-20 μm in the *ab* plane. The density of textured Bi2223 is equal to 5.3 g/cm$^3$.

Samples for magnetic measurements had cubic form with the edges of 1.6 mm. The $M(H)$ dependencies were measured at temperatures of 4.2, 14, 25, 40, 60 and 80 K. The textured samples were investigated in two orientations of field $\boldsymbol{H}$ relative to the predominant direction of the crystallites in the textured sample: the magnetic field is directed perpendicularly to the texturing plane ($\boldsymbol{H} \parallel \boldsymbol{c}$) and along the texturing plane ($\boldsymbol{H} \parallel \boldsymbol{ab}$).

Figure 2 demonstrates the magnetization loops of the porous Bi2223 measured at temperatures ($T$) of 4.2 K - 80 K. Figure 3 demonstrates the magnetization loops of the textured Bi2223 measured at temperatures of 4.2 K - 80 K for $\boldsymbol{H} \parallel \boldsymbol{c}$ and $\boldsymbol{H} \parallel \boldsymbol{ab}$. For the $M(H)$ dependencies at $T$ = 40, 60 and 80 K, values of the irreversibility field $H_{\mathrm{irr}}(T)$ are smaller than the maximal applied field $H_{\mathrm{m}}$ = 60 kOe. At any fixed temperature, $H_{\mathrm{irr}}$ of the porous Bi2223 is smaller than the irreversibility fields of the textured Bi2223 and for the textured Bi2223 $H_{\mathrm{irr}}$ for $\boldsymbol{H} \parallel \boldsymbol{c}$ is smaller than $H_{\mathrm{irr}}$ for $\boldsymbol{H} \parallel \boldsymbol{ab}$. The $M(H)$ dependencies of the textured Bi2223 are scaled according to $M^{H\parallel c}(H) = \gamma M^{H\parallel ab}(\gamma H)$ [9] with the anisotropy coefficient $\gamma$ = 2.4. Results of the detailed investigation of anisotropy of the texture will be presented in the other work.

The experimental $M(H)$ dependencies demonstrate that the asymmetry increases with temperature and field. At $T$ = 4.2 K the magnetization loops are nearly symmetric relative to the $M$ = 0 axis at fields of 10-55 kOe, but at higher temperatures the magnetization loops are distinctly asymmetric. At $T$ = 80 K the $M(H)$ dependencies are reversible for the porous Bi2223 at fields higher than 1.5 kOe and for the textured Bi2223 at fields higher than 2 kOe for $\boldsymbol{H} \parallel \boldsymbol{c}$ and 4 kOe for $\boldsymbol{H} \parallel \boldsymbol{ab}$.

## 4. Discussion

The experimental data were fitted in the frame of the developed model. The main fitting parameters are the depth of surface equilibrium layer $l_s$ and the product of $j_{\mathrm{c}}$ and the average granule radius $R$. We used the simple dependence $l_s(H) = l_s(H = 0)\,(1 + H/(0.3H_{\mathrm{irr}}))$ that works good for all described curves at any $T$. The calculated magnetization loops depends very slightly on different distribution functions $\varphi(R)$ for both the textured Bi2223 and the porous Bi2223 such that we disregarded a dispersion of the granule parameters and used average values.

In figures 2 and 3 the computed curves are displayed together with the experimental loops. There is the good agreement between the presented data at all temperatures. Some discrepancy may be due to a demagnetizing factor of the sample and grains [10]. The value of the fitted $l_s$ at $H = 0$ grows from $l_s = 200$ nm at $T = 4.2$ K to $l_s \sim 3000$ nm at $T = 80$ K (see Fig. 4a). For all samples, at $H = 40$ kOe the average radius $R$ is smaller than values of $l_s$ at low temperatures but $l_s$ becomes greater than $R$ at $T$ above 40 K (Fig. 4b). Figure 5 shows the intragranular critical current density estimated from the fit with $R = 10$ μm for textured Bi2223 and $R = 20$ μm for the porous Bi2223.

In conclusion, the magnetization loops of porous and textured Bi2223 were measured in fields up to 60 kOe at different temperatures. The $M(H)$ dependence asymmetry increases with increasing temperature. To describe the asymmetric magnetization loops, the extended variant of the critical state model was developed. In the presented model, the asymmetry is referred to the single parameter $l_s$, which determines the position of the boundary between the surface layer with the thermal equilibrium magnetization, and the central region with the critical state magnetization. As following from the model proposed, the $M(H)$ dependence of the sample consisting of sufficiently small superconducting granules (~λ) should be fully reversible at all temperatures.

**Acknowledgements**

D.M. Gokhfeld is thankful to S.A. Satzuk and Yu.S. Gokhfeld for discussions. The work is supported by project № 7 of RAS Program “Quantum physics of condensed matter”, Federal Target Program "Research and scientific-pedagogical cadres of Innovative Russia" and grant № 13 of Lavrentyev competition of young researchers of SB RAS.

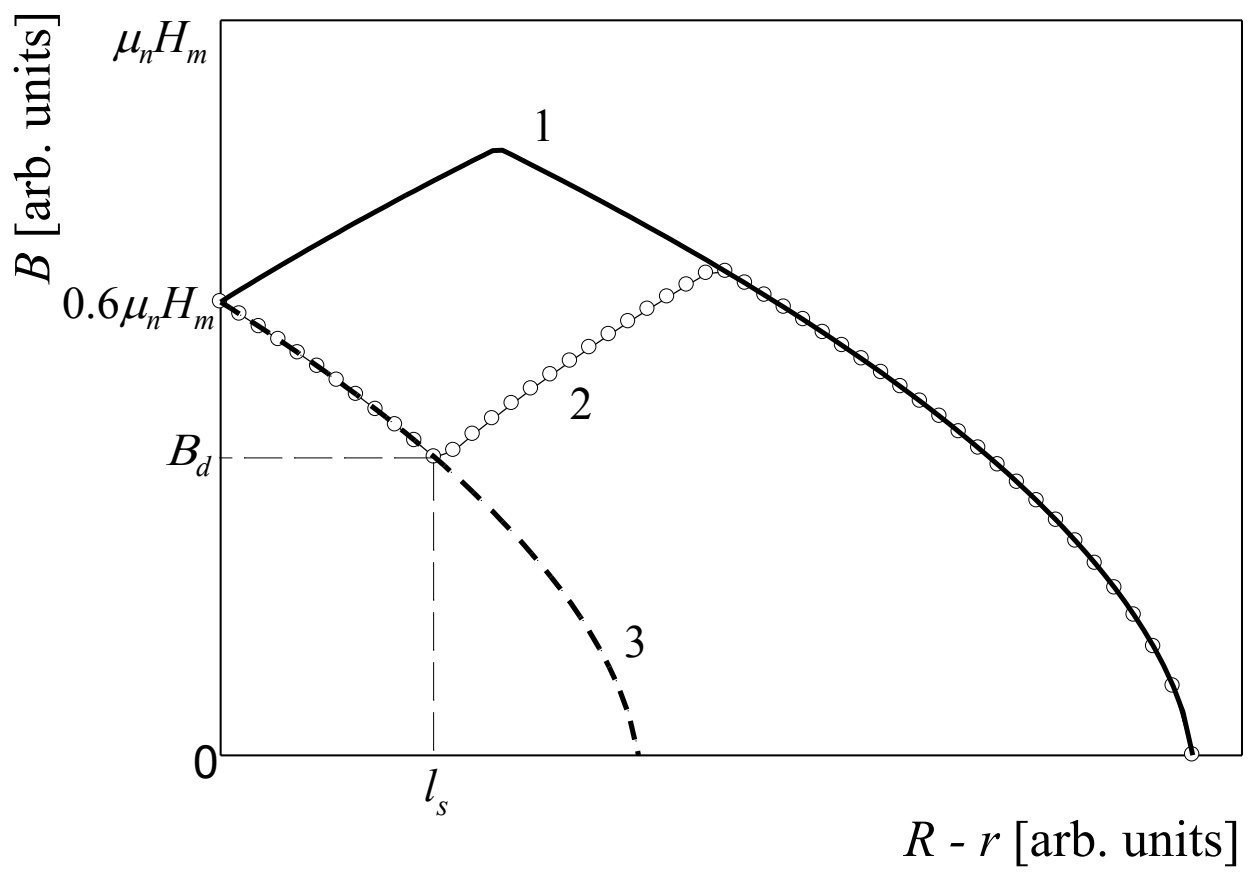

**Fig. 1.** Dependence of induction $B$ on depth $R - r$ from the granule surface after field decreased from $H_m$ to 0.6 $H_m$ for $l_s = 0$ (curve 1, solid line), $0 \le l_s \le R$ (curve 2, circles), $l_s = R$ (curve 3, dotted line).

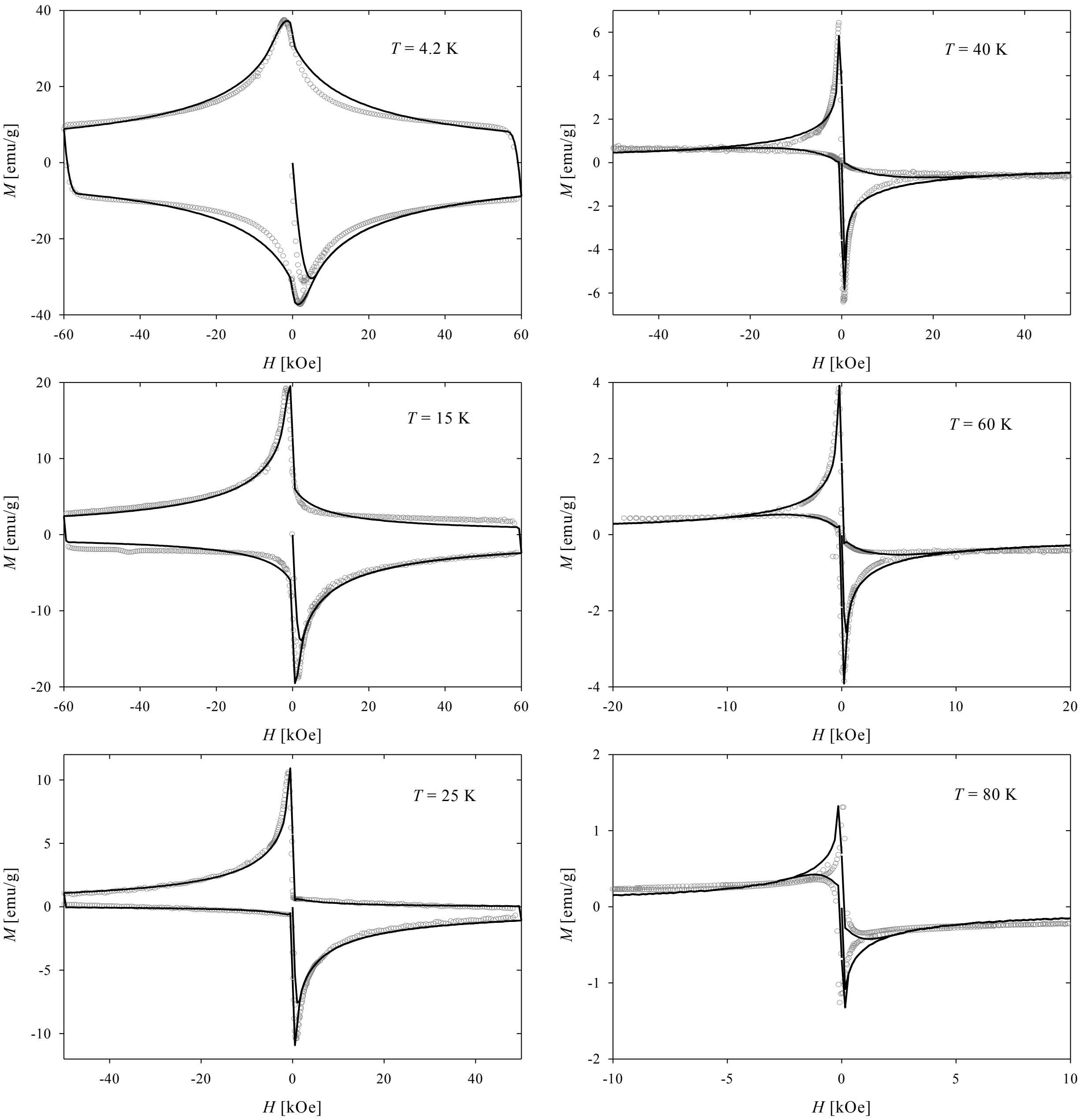

**Fig. 2.** Dependencies of magnetization $M$ on magnetic field $H$ of the porous Bi2223. Experimental data (symbols) and computed curves (solid lines).

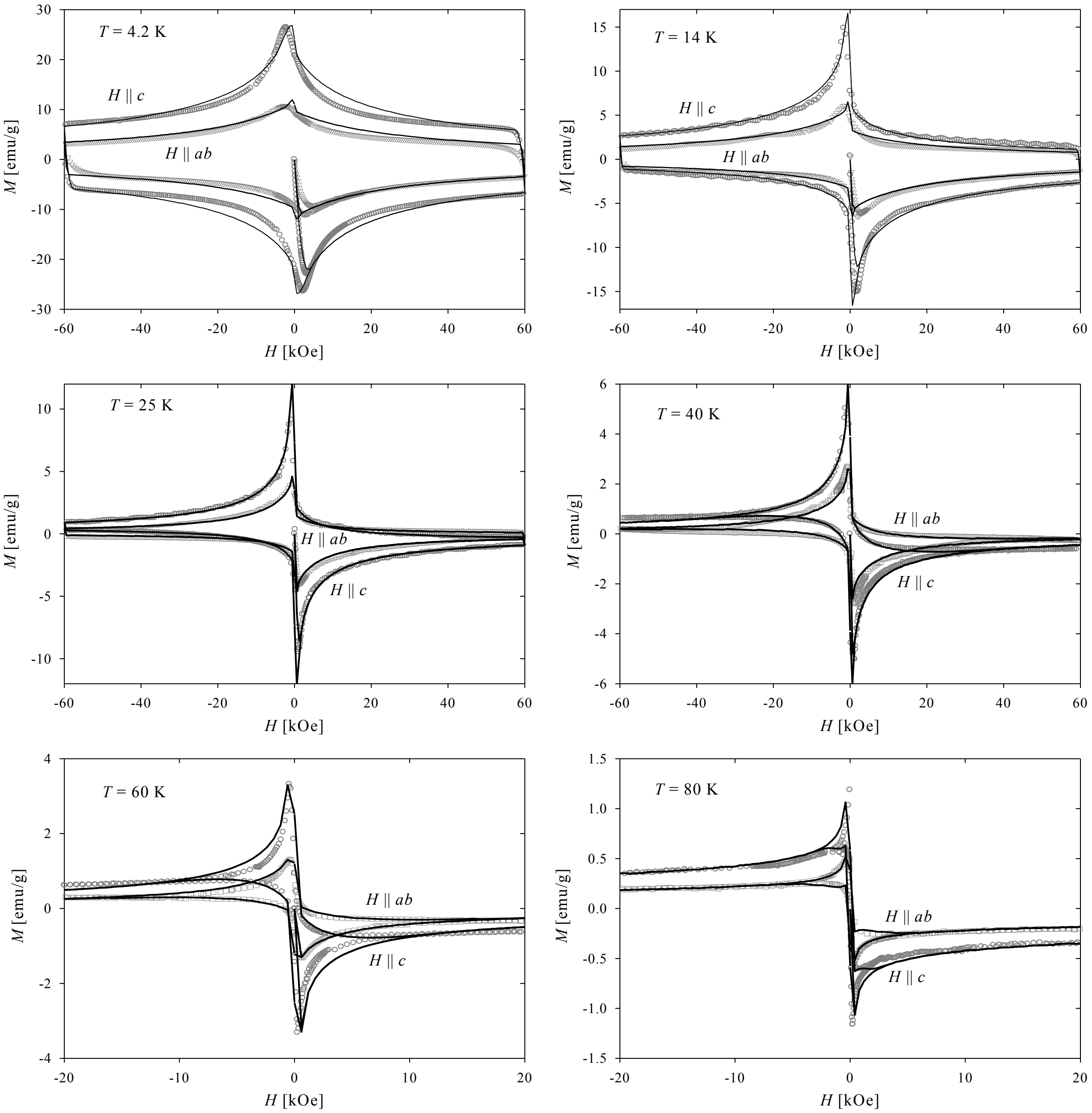

**Fig. 3.** Dependencies of magnetization *M* on magnetic field *H* of the textured Bi2223. Experimental data (symbols) and computed curves (solid lines).

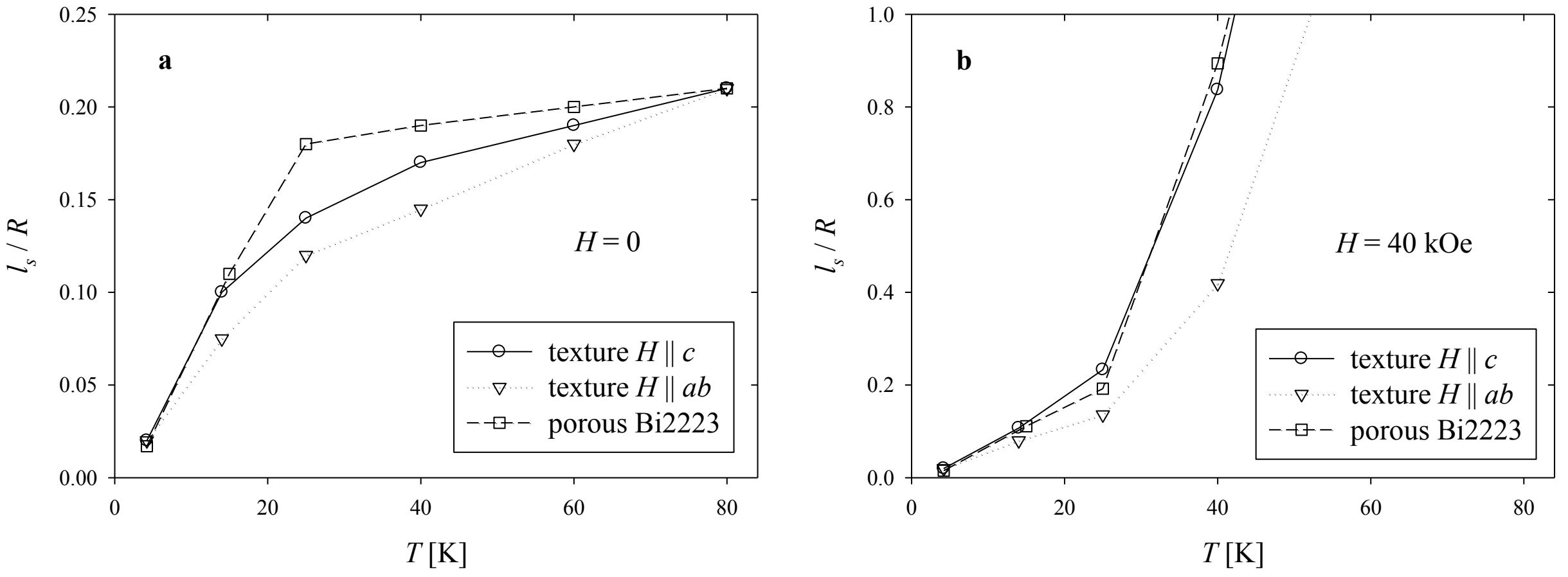

**Fig. 4.** Dependence of boundary depth $l_s$, divided by the average radius of granule $R$, on temperature at $H = 0$ (a) and $H = 40$ kOe (b).

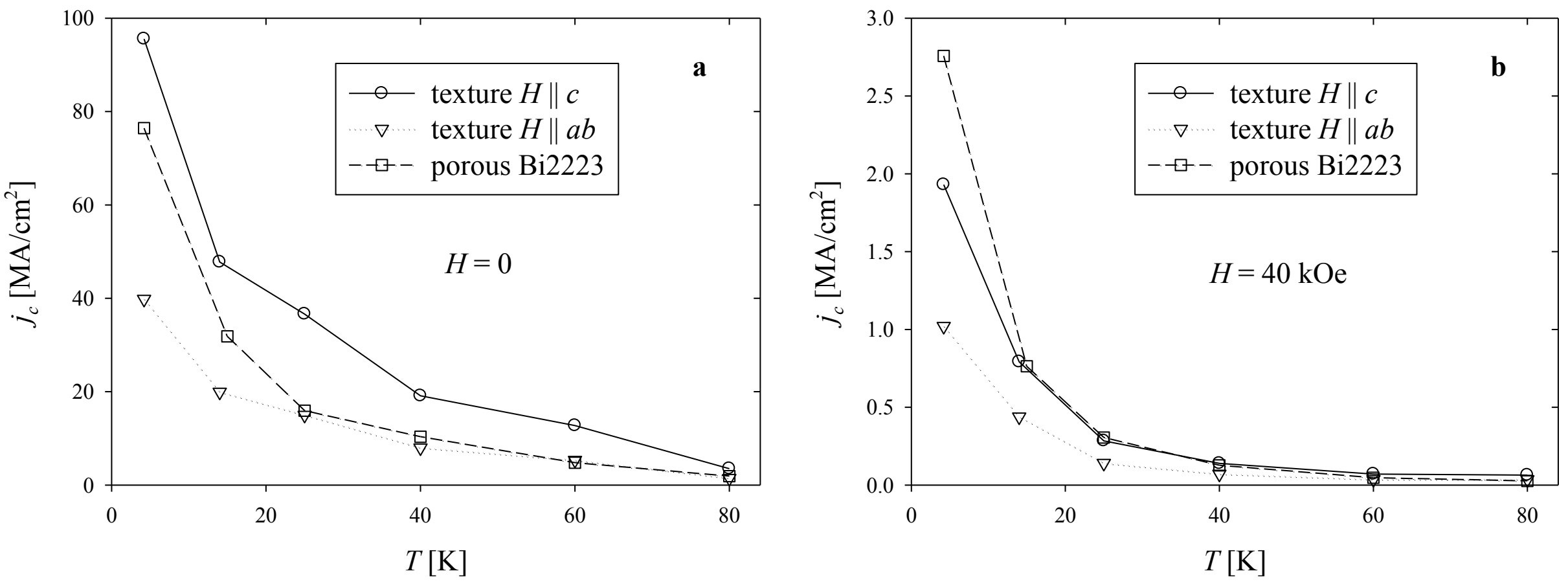

**Fig. 5.** Critical current density $j_c$ versus temperature at $H = 0$ (a) and $H = 40$ kOe (b).